\documentclass[%
 reprint,
 amsmath,amssymb,
 aps,
 prb,
]{revtex4-2}

\usepackage{braket}
\usepackage{graphicx}
\usepackage{dcolumn}
\usepackage{bm}
\usepackage{amsmath}
\usepackage[colorlinks,allcolors=blue]{hyperref}

\def \samplex {Cr$_y$(Bi$_{1-x}$Sb$_x$)$_{2-y}$Te$_3$ \hspace{1 pt}}
\def \sample {Cr$_{0.1}$(Bi$_{0.2}$Sb$_{0.8}$)$_{1.9}$Te$_3$ \hspace{1 pt}}

\begin{document}

\title{Phonon and defect mediated quantum anomalous Hall insulator to metal transition in magnetically doped topological insulators}
\author{Akiyoshi Park$^{1,2}$, Adrian Llanos$^2$, Chun-I Lu$^1$, Yinan Chen$^{1,2}$, Sebastien N. Abadi$^1$,  \\
Chien- Chang Chen$^{1,2}$, Marcus L. Teague$^{1,2}$, Lixuan Tai$^3$, Peng Zhang$^3$, Kang L. Wang$^3$, and Nai-Chang Yeh$^{1,2}$  }
\altaffiliation{Corresponding Author: \href{mailto:ncyeh@caltech.edu}{ncyeh@caltech.edu} }
\affiliation{$^1$ Department of Physics, California Institute of Technology, Pasadena, CA 91125, USA}
\affiliation{$^2$ Institute for Quantum Information and Matter, California Institute of Technology, Pasadena, CA 91125, USA}
\affiliation{$^3$ Department of Electrical Engineering, University of California, Los Angeles, CA 90095, USA}

\date{\today}

\begin{abstract}
Quantum Anomalous Hall (QAH) state in six quintuple layer \sample thin films were studied through scanning tunneling spectroscopy (STS) and electrical transport measurements. While the surface state is gapless above the Curie temperature ($T_\mathrm{C} \approx 30$ K), scanning tunneling spectroscopy (STS) of the sample reveals a topologically non-trivial gap with an average value of $\approx 13.5$ meV at 4.2 K below the ferromagnetic transition. Nonetheless, areal STS scans of the magnetic topological insulator exhibit energy modulations on the order of several meV's in the surface bands which result in the valence band maximum in some regions becoming higher than the energy of the conduction band minimum of some other regions that are spatially separated by no more than 3 nm. First principle calculations demonstrate that the origin of the observed inhomogeneous energy band alignment is an outcome of many-body interactions, namely electron-defect interactions and electron-phonon interactions. Defects play the role of locally modifying the energy landscape of surface bands while electron-phonon interactions renormalize the surface bands such that the surface gap becomes reduced by more than 1 meV as temperature is raised from 0 to 4.2 K. These many-body interactions at a finite temperature result in substantial increase of electron tunneling across the spatially separated conduction band pockets even for finite temperatures well below $T_\mathrm{C}$ , thus driving the magnetic topological insulator out of its QAH insulating phase into a metallic phase at a relatively low temperature.
\end{abstract}

\maketitle

\subsection{Introduction}

Experimental observations of the quantum anomalous Hall effect (QAHE) amongst Cr-doped (Bi,Sb)$_2$Te$_3$, a system of magnetic topological insulators (MTIs), have been bound to below sub-Kelvin temperatures \cite{doi:10.1126/science.1234414, PhysRevLett.113.137201, Checkelsky2014}. Such constraint of the QAHE to low temperatures is unexpected given that these MTIs exhibit long-range ferromagnetic ordering at much higher temperatures \cite{doi:10.1063/1.3688043}. Specifically, transport measurements of Cr-doped (Bi,Sb)$_2$Te$_3$ have shown bulk Curie temperatures ($T_\mathrm{C}$) ranging from 20 to 30 K, whereas the appearance of QAHE has been limited to temperatures below 0.1 K \cite{doi:10.1063/1.3688043}.

It has been proclaimed that the bottle neck in the temperature range of the QAHE is the lack of spatially homogeneous ferromagnetic ordering such that the region with gapped topological surface states is beneath the percolation threshold at higher temperatures \cite{PhysRevB.92.201304, https://doi.org/10.1002/adma.201600919, Lee1316, doi:10.1126/sciadv.1500740} . Nonetheless, significantly improved magnetic homogeneity using either magnetic modulation doping in heterostructures \cite{10.1063/1.4935075} or co-doping Cr with V atoms \cite{https://doi.org/10.1002/adma.201703062} could only raise the threshold temperature for QAHE to approximately 2 K. 

Fundamentally for QAHE to ensue, the effective Zeeman field emanating from ferromagnetic ordering closes the topologically trivial hybridization gap through band inversion and opens up a non-trivial magnetic gap, thus creating a Chern insulator phase characterized by the topological invariant, $C =\pm 1$, referred to as the Chern number. Although some experiments have reported the presence of the magnetic gap opening in the surface states of MTIs \cite{Lee1316, PhysRevLett.112.056801, Chen_2015, 10.1063/1.4990548, doi:10.1126/science.1189924}, there are contradictory results that indicate otherwise \cite{10.1063/5.0039059, 10.1063/5.0070557}. There are also reports of opening of a topological non-trivial gap being masked by mid-gapped impurity states \cite{Sessi2016, PhysRevB.91.201411} or bulk states \cite{PhysRevLett.115.057206}. In addition, it has been argued that the Dirac point in MTIs is degenerate with the bulk valence band, thereby compromising the conductivity \cite{Li2016, doi:10.1126/sciadv.aaz3595}. Hence, the issue regarding whether magnetic doped (Bi,Sb)$_2$Te$_3$ are true topological insulators has not been settled.
\!
\\
\indent In particular, the lack of general agreement amongst reports concerning the electronic structure of these materials is attributed to inadequate distinctions between MTI samples in the bulk and the two-dimensional (2D) limit. The two are fundamentally different, given that the QAHE has only been observed in epitaxial films \cite{https://doi.org/10.1002/adma.201600919, Kou2015, doi:10.1126/sciadv.aaz3595}. The absence of quantized transport in bulk single crystal is attributed to higher defect concentration such that the chemical potential falls outside the surface gap \cite{Lee1316, doi:10.1021/acs.nanolett.0c02873}, and a larger structural inversion asymmetry that appears to be a obstacle for a topological phase transition to the Chern insulator state \cite{PhysRevLett.111.146802}. The lack of a consensus thus calls for a more thorough investigation to map out the energies of the surface and bulk electronic states above and below the topological phase transition temperature in 2D MTI-epitaxial films, which duly exhibit a QAH phase.

\subsection{Experimental and Computational Methods}
\subsubsection{Sample growth}
\samplex thin films with an average thickness of 6 quintuple layers (QLs) were grown by molecular beam epitaxy (MBE) on (111) GaAs substrates in an ultrahigh vacuum Perkin-Elmer instrument. The details of the method have been outlined in Ref. \cite{doi:10.1126/sciadv.aaz3595}. The thickness of the sample is determined through the oscillation of Reflection High Energy Electron Diffraction (RHEED) patterns over the growth process. 

Although there are terrace-like structures and thickness variation over different regions of on the thin film (Fig. \ref{topography}(a)), the average thickness is determined to be 6 QLs, which is sufficiently thick to suppress the hybridization gap between the bottom and top surface states to $\approx 5$ meV, but still thin enough to reduce the effect of structural inversion asymmetry, that prevents band inversion to a topological non-trivial phase \cite{refId0, Zhang2010}.

\subsubsection{Scanning tunneling microscopy/spectroscopy}
Scanning tunneling microscopy/spectroscopy (STM/STS) measurements were performed in a homemade microscope connected to a RHK R9 controller. To achieve liquid helium temperatures, small traces of ultrahigh purity He ($99.999\%$) exchange gas were added to the vacuum chamber. Tunneling spectroscopy was measured through an AC lock-in amplifier with a bias modulation of 10 mV amplitude. Mechanically cleaved Pt-Ir alloy wires were used as the STM tips.

\subsubsection{Transport measurements}
Thin film samples for transport measurements were first patterned into a Hall bar geometry. Temperature and magnetic field dependence of longitudinal and Hall resistance was measured by using a Physical Property Measurement System (PPMS) by Quantum Design. Three SRS SR830 Lock-in Amplifiers were employed: two were used to monitor the Hall voltage and the longitudinal voltage, and one was used to supply a current by applying an AC voltage of $V_\mathrm{rms} = 10$ mV and frequency between $f=5-15$ Hz across a 1 M$\Omega$ reference resistor. 

\subsubsection{First principle calculations}
Density functional theory (DFT) calculations were performed by Quantum Espresso (QE) using a fully relativistic pseudopotential and setting up non-collinear calculations to incorporate the effects of spin-orbit coupling \cite{Giannozzi_2009, Giannozzi_2017}. For calculating the electronic structure of the parent compound Sb$_2$Te$_3$, initially a ($4\times4\times1$) k-point mesh was constructed to obtain self-consistent electron densities through self-consistent field (SCF) calculations followed by non-self-consistent field (NCF) calculations to obtain eigenvalues of a finer ($40\times40\times1$) k-point mesh.

Furthermore, to implement and investigate various perturbative disorder to topological insulators such as defects, magnetic dopants, and electron-phonon coupling, which break translational lattice symmetry, a change from reciprocal space to real space basis is required. This was performed through using Bloch states obtained from QE to construct a real-space lattice Hamiltonian by computing Maximally Localized Wannier Functions (MLWFs) via the Wannier90 package \cite{MOSTOFI20142309}.  For each atom in the unit cell, initial projections were represented as $p_x$, $p_y$ and $p_z$ orbitals with spin-up and -down, given that the $p$-oribtals of (Bi,Sb) and Te are the most relevant bands lying close to the Fermi level ($E_\mathrm{F}$) \cite{Zhang2009}. Hence, a 3 QL unit cell consisting of 30 atoms would yield a $(90\times90)$-matrix Hamiltonian, and a $(180\times180)$-matrix Hamiltonian for a 6 QL unit cell.

The electronic structures of the doped system were calculated through construction of supercells with various doping configurations. The supercell bandstructures were determined through SCF and NCF calculations using QE followed by Wannerisation, as with the case for the unit cell calculations. Since the Brillouin zone (BZ) of the supercell is smaller than the BZ of the unit cell, an extra step of band unfolding was necessitated, which was done through the Wanniertools post processing software \cite{WU2018405}.

To realize the effects of finite temperatures, electron-phonon coupling calculations were performed through a density functional perturbation theory (DFPT) approach.  Dynamical matrices were computed using the phonon module in QE for a ($3\times3\times1$) q-point mesh. The electron-phonon coupling parameters were then calculated up to the linear response regime through the Electron-phonon Wannier (EPW) code using the dynamical matrices computed from QE \cite{PONCE2016116}.

\subsection{Results}

\begin{figure}[t]
\includegraphics[width =0.5\textwidth]{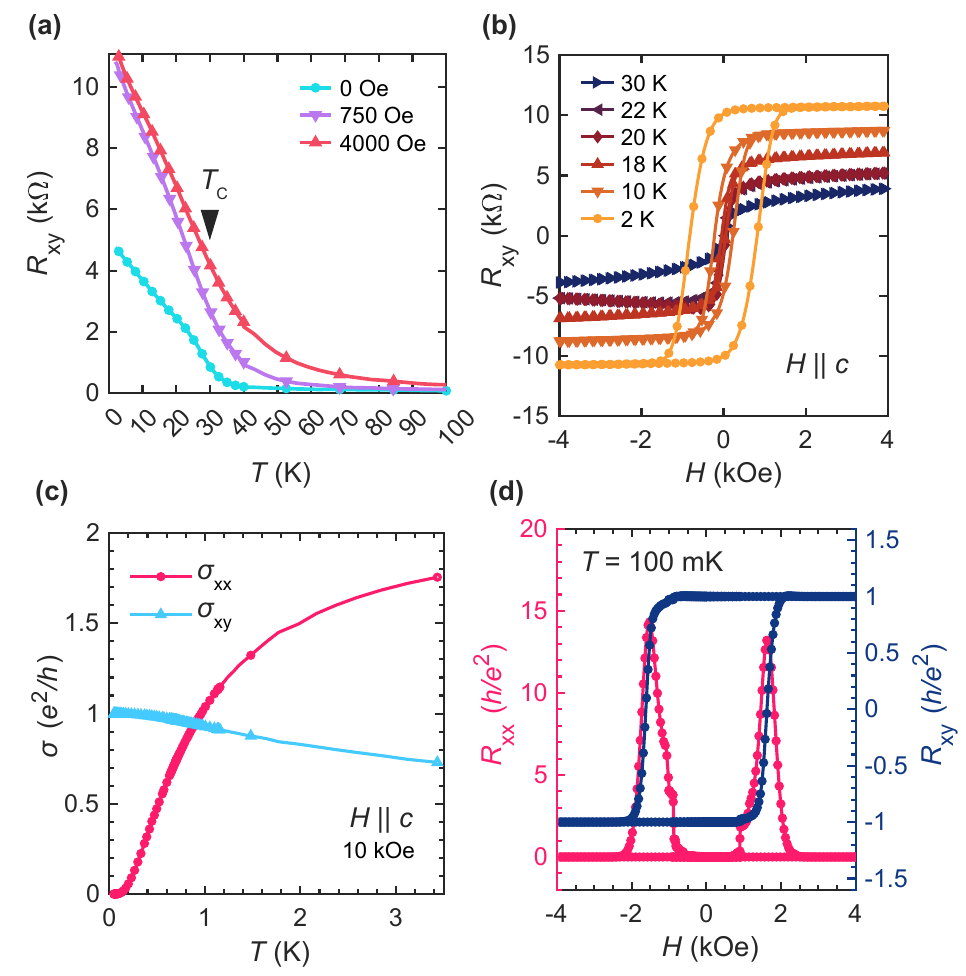}
\caption{\label{transport} Temperature and magnetic field dependence of longitudinal and Hall Resistance in \sample (a) The temperature dependence of $R_\mathrm{xy}$ measured at various magnetic fields. A rapid increase in zero-field $R_\mathrm{xy}$ appears around $T = 30$ K, suggesting the onset of ferromagnetism. (b) Hysteric behavior in $R_\mathrm{xy}$ dependence of magnetic field ($H$) at different temperatures between 30 K and 2 K.(c)  $\sigma_\mathrm{xy}$ and $\sigma_\mathrm{xx}$ upon field cooling with $H = 10$ kOe exhibiting QAHE at $T = 100$ mK. (d) Magnetic field sweep  of $R_\mathrm{xy}$ and $R_\mathrm{xx}$ at $T = 100$ mK.}
\end{figure}

\subsubsection{Sample Characterization}
6 QL \samplex thin films investigated in this work consist of a Cr-doping level of $y=0.10$, which is the optimal doping for ferromagnetic order \cite{doi:10.1126/sciadv.aaz3595}, and a Sb-doping level of $x=0.80$ to tune the Fermi energy ($E_\mathrm{F}$) right at the Dirac point ($E_\mathrm{D}$) \cite{doi:10.1126/science.1234414, Zhang2011}. The presence of bulk ferromagnetic order due to Cr-doping is manifested by the emergence of a finite Hall resistance ($R_\mathrm{xy}$) across the sample upon zero-field cooling below $T_\mathrm{C}$ at approximately 30 K (Fig. \ref{transport}(a)), indicative of time-reversal symmetry breaking. The bulk ferromagnetism below $T_\mathrm{C}$ is further verified by the opening of a hysteresis loop in the field dependent Hall resistance data (Fig. \ref{transport}(b)), which reveals a finite coercive field ($H_\mathrm{C}$) that decreases with increasing temperature and diminishes to zero at approximately 30 K. At an even lower temperature of $T=100$ mK, $\sigma_\mathrm{xy}$ reaches a quantized value of $e^2/h$, while the longitudinal resistivity ($\sigma_\mathrm{xx}$) is suppressed to zero (Fig. \ref{transport}(c,d)), verifying that the sample is indeed a QAH-insulator. 

\begin{figure}[t]
\includegraphics[width =0.5\textwidth]{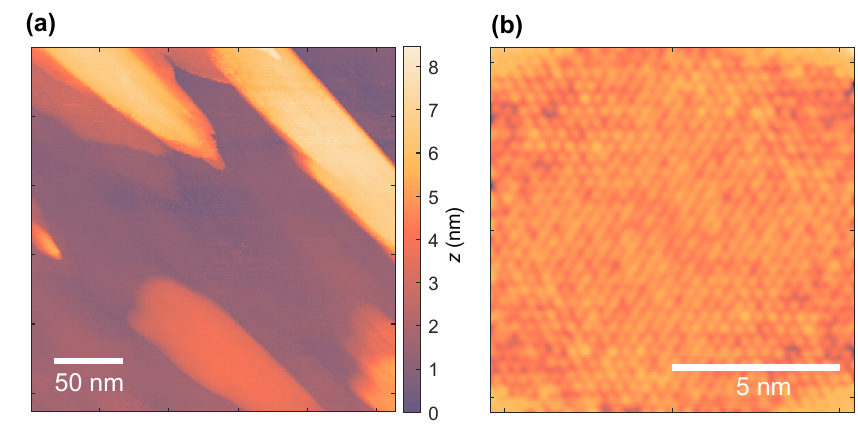}
\caption{\label{topography} STM Topographic images at $T = 4.2$ K. (a) Images of terrace like structures were observed over a larger area of (300 nm x 300 nm) due to the layered nature of \sample, obtained with scanning parameters of $I_\mathrm{t} = 100 $ pA and $V_\mathrm{b} = 1 $ V. (b) A smaller area of (11 nm x 11 nm),  where individual atoms were resolved with scanning parameter of $I_\mathrm{t} = 1 $ nA and  $V_\mathrm{b} = 100 $ mV. The atomic scale image is processed through Fourier filtering to improve the signal-to-noise ratio of the image.}
\end{figure}

Large area STM topographic images indicate that the sample surface is atomically flat, exhibiting islands with terrace-like features (Fig. \ref{topography}(a)), whereas atomically resolved images in smaller areas reveal a triangular lattice pattern, representative of the crystalline structure of (Bi,Sb)$_2$Te$_3$ in the $ab$-plane. Noticeably, the unevenness in the topography over a nm-scale represented by the darker color patches in Fig. \ref{topography}(b) may be attributed to collective contributions from Bi/Sb alloying \cite{PhysRevB.97.125150}, Cr-dopants (with an atomic radius much smaller than those of Bi and Sb atoms), or dilute crystalline defects, including interstitial and vacancies.

To understand the change in the properties of the samples across the ferromagnetic transition, where time-reversal symmetry is broken, scanning tunneling spectroscopy (STS) was performed both above $T_\mathrm{C}$ at liquid nitrogen temperature (77 K) and below $T_\mathrm{C}$ at liquid helium temperature (4.2 K).

\begin{figure}[t]
\includegraphics[width =0.5\textwidth]{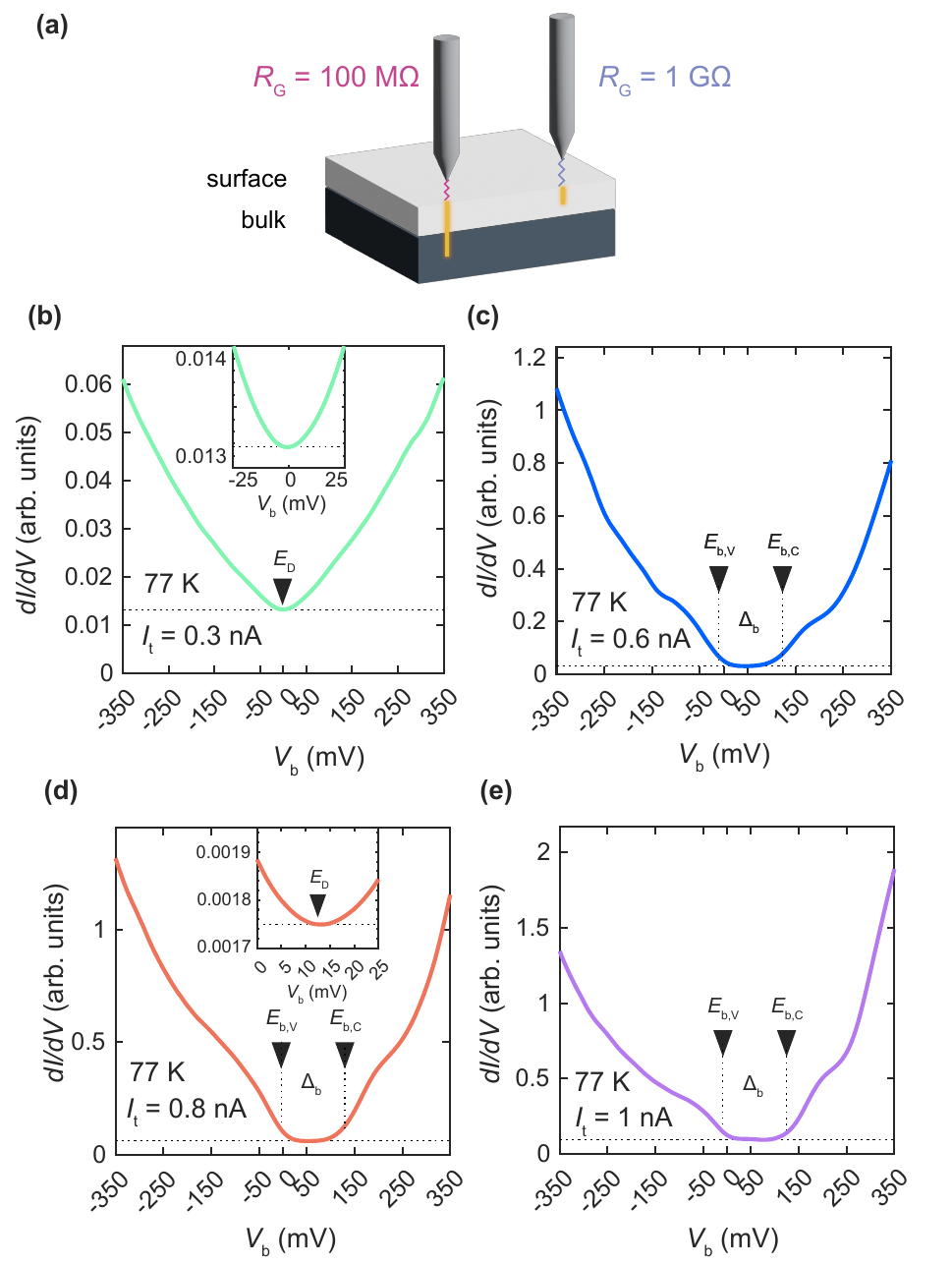}
\caption{\label{STS77K}  Scanning tunneling spectra taken with different tunneling resistance at $T = 77$ K. (a) Schematic representation of tip-sample distance changing as a consequence of changing the tunneling resistance from 100 M$\Omega$ to 1 G$\Omega$. The higher the tunneling current, the closer the tip-sample distance in which the point spectra were measured. (b-e) Tunneling resistance dependent point spectra were obtained with a fixed bias of $V_\mathrm{b} = 100$ mV and variable tunneling currents of $I_\mathrm{t} = 300$ pA, $I_\mathrm{t} = 600$ pA,  $I_\mathrm{t} = 800$ pA, $I_\mathrm{t} = 1000$ pA. The horizontal dotted lines are the baselines which indicate the minima of each spectrum. The insets exhibit the low energy region of the spectra with the corresponding Dirac point.}
\end{figure}

\subsubsection{STM investigation above the Curie temperature}

\begin{figure}[t]
\includegraphics[width =0.5\textwidth]{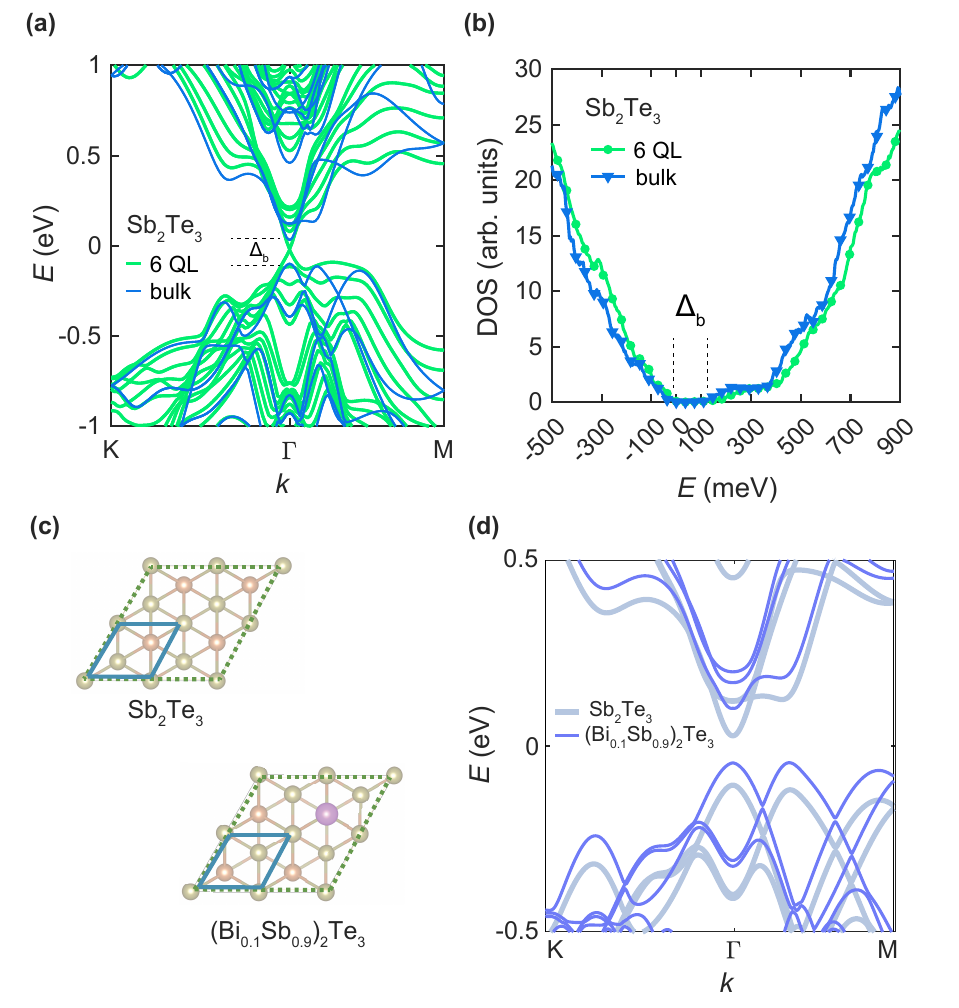}
\caption{\label{bandstructure}  Calculated electronic structure of (Bi,Sb)$_2$Te$_3$ from DFT. (a) The band structure and (b) the DOS of Sb$_2$Te$_3$ with a bulk configuration and a 6 QL slab configuration. (c) Crystalline $ab$-plane of a 2$\times$2$\times$1 supercell for the parent compound, Sb$_2$Te$_3$, and the doped compound, (Bi$_{0.1}$Sb$_{0.9}$)$_2$Te$_3$. The dashed lines represent the supercell and the solid lines represent the unit cell. (d) The unfolded band structure of both Bi-doped and parent bulk supercell structure. }
\end{figure}

Information associated with the local density of states (LDOS) is acquired by conducting differential conductance ($dI/dV$) spectroscopy, in which the height of the STM tip relative to the sample surface is kept constant while measuring the tunneling current ($I_\mathrm{t}$) as a function of the bias voltage ($V_\mathrm{b}$) between the sample and the tip with a lock-in amplifier. The spatial separation between the tip and the sample is determined by the tunneling resistance ($R_\mathrm{G} = V_\mathrm{b} / I_\mathrm{t}$), where a higher $R_\mathrm{G}$ corresponds to a larger tip-sample separation (Fig. \ref{STS77K} (a)). Figures \ref{STS77K} (b-e) exhibit the tunneling spectra obtained from an atomically flat surface area at 77 K with varying $R_\mathrm{G}$. With $R_\mathrm{G}$ = 333 M$\Omega$ ($I_\mathrm{t}$ = 300 pA, $V_\mathrm{b}$ = 100 mV), a V-shaped energy dependence was evident from the $dI/dV$ spectrum, which was representative of the Dirac surface states with the minimum $dI/dV$ appearing at the $E_\mathrm{D}$ (Fig. \ref{STS77K} (b)), and the finite LDOS at and around $E_\mathrm{D}$ indicates a virtually gap-less surface state as expected above $T_\mathrm{C}$. Remarkably, $E_\mathrm{D}$ nearly coincides with the Fermi level (corresponding to $E = 0$ in the tunneling spectrum) \cite{Zhang2011}, which is in stark contrast to most single crystal samples with $E_\mathrm{D}$ more than 100 meV above $E_\mathrm{F}$ \cite{PhysRevLett.111.176802, Lee1316, doi:10.1021/acs.nanolett.0c02873}.

As $R_\mathrm{G}$ is reduced such that the STM tip is brought closer to the sample surface, the $dI/dV$ curve loses its V-shaped spectrum and exhibits kink-like features (Fig.  \ref{STS77K}(c-e)). Particularly, at low energies nearby $E_\mathrm{F}$, an energy region of vanishing LDOS is recognizable. This energy region of vanishing LDOS corresponds to the bulk bandgap ($\Delta_\mathrm{b}$) \cite{Zhang2009}, as suggested by band structure obtained by first principle calculations (Fig. \ref{bandstructure}). When reducing the tunneling resistance, the STM tip is sufficiently close to the sample surface, such that electron tunneling is no longer restricted exclusively between the tip and the surface states, but also between the tip and the bulk states that are only a few atomic layers below the surface. Tuning $R_\mathrm{G}$, accordingly,  allows selectively probing either/both the surface or/and bulk states in the \sample thin films of our investigation \cite{PhysRevLett.108.066809}. 

Notably, the Dirac point is located within $\Delta_\mathrm{b}$ (Fig. \ref{STS77K}(d)), and the value of $\Delta_\mathrm{b}$ quantified through STS is on the same energy scale as that calculated from first principles for the parent compound, Sb$_2$Te$_3$ as well as that of the doped (Bi$_{0.1}$Sb$_{0.9}$)$_{2}$Te$_{3}$ compound (Fig. \ref{bandstructure}(d)). Comparing the bulk band structure of Sb$_2$Te$_3$ and (Bi$_{0.1}$Sb$_{0.9}$)$_{2}$Te$_{3}$, we note that the $E_\mathrm{F}$ value of the Bi-doped bulk band structure is $70$ meV higher than the $E_\mathrm{F}$ of the bulk band structure of the parent compound, owing to electron donation from the Bi atoms. Otherwise, the bulk band structures of the two compounds are nearly identical, with a consistent $\Delta_\mathrm{b}$.

Moreover, we address the issue of the absence of spectral evidence for the edge states near the step edges shown in Fig. \ref{topography}(a). We attribute the lack of spectral characteristics of the edge state to  local wavefunction hybridization between the surface states above and below the step edge, similar to the finite size effect encountered in the ultrathin film ($<$ 6 QL) limit, which hinders the manifestation of spectral characteristics of the edge states at a step edge.

\begin{figure}[t]
\includegraphics[width =0.5\textwidth]{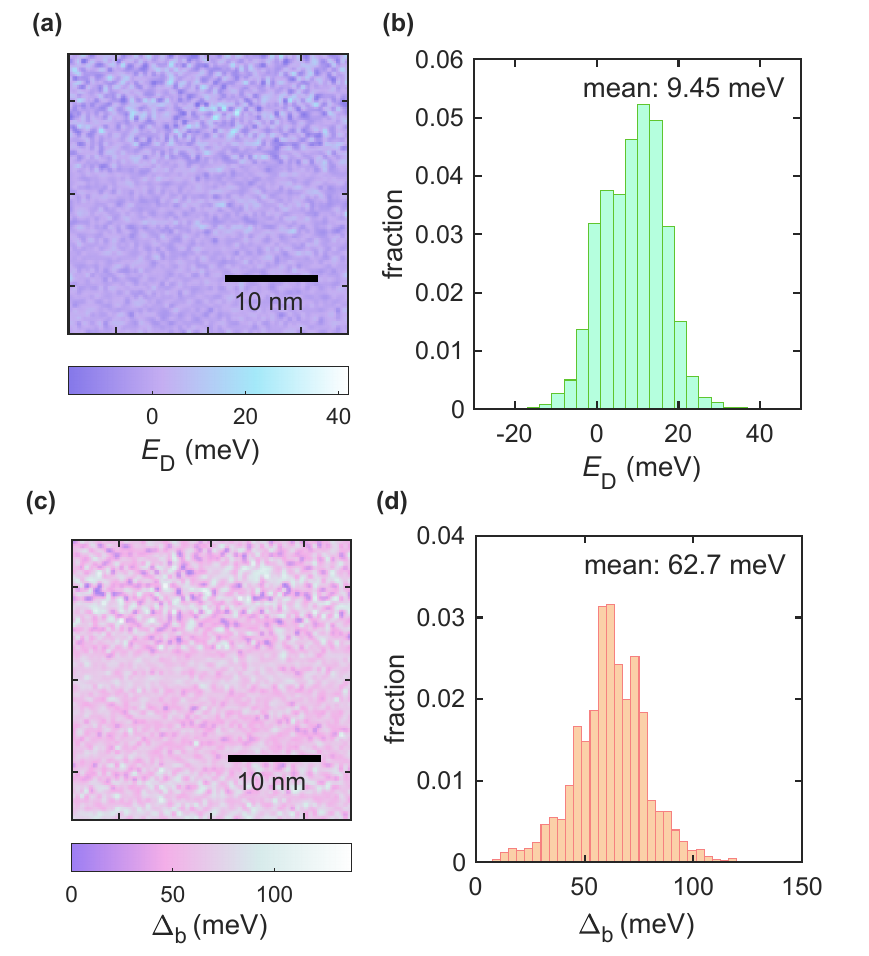}
\caption{\label{stsmap}  Spatial distribution of $E_\mathrm{D}(\vec{r})$ and $\Delta_\mathrm{b}(\vec{r})$ at 77 K. (a) The spatial map of $E_\mathrm{D}(\vec{r})$ measured at 77 K, (b) with the corresponding histogram. (c) The spatial map of $\Delta_\mathrm{b}(\vec{r})$ measured at 77 K, (d) with the corresponding histogram.}
\end{figure}

To evaluate the spatial distribution of $E_\mathrm{D}(\vec{r})$ and $\Delta_\mathrm{b}(\vec{r})$, the tunneling resistance was tuned to 167 M$\Omega$ ($I_\mathrm{t}$ = 600 pA, $V_\mathrm{b}$ = 100 mV), such that information from both the surface and the bulk are probed. With this fixed  resistance, the $dI/dV$ spectra were taken over a (30 nm $\times$ 30 nm) area. The resulting distributions are depicted in Fig. \ref{stsmap}. The average Dirac point lies at  $\braket{E_\mathrm{D}(\vec{r})}=(9.45 \pm 7.52)$ meV, where the uncertainty here and henceforth represents the first standard deviation of the spatial variation. In this nm-scale scanned area,  $E_\mathrm{D}(\vec{r})$ has a narrow distribution, tightly bound to around $E_\mathrm{F}$ (Fig. \ref{stsmap}(a,b)), further signifying how well the Bi/Sb doping was tuned in our samples to its ideal energy location to optimize the QAHE. In addition, the magnitude of the mean bulk band gap was at $\braket{\Delta_\mathrm{b}(\vec{r})} =(62.7\pm 16.5)$ meV (Fig. \ref{stsmap}(c,d)), with the mean energies of the bulk valence band maxima and the bulk conduction band minima being $\braket{E_\mathrm{b,V}(\vec{r})}=(-13.6 \pm 16.5)$ meV and $\braket{E_\mathrm{b,C}(\vec{r})}=(48.8 \pm 12.9)$ meV, respectively. The spatial distribution maps of Fig. \ref{stsmap}(a,c) reaffirm the statistical homogeneity on the nanometer scale.

\subsubsection{STM investigation below the Curie temperature}

\begin{figure}[t]
\includegraphics[width =0.5\textwidth]{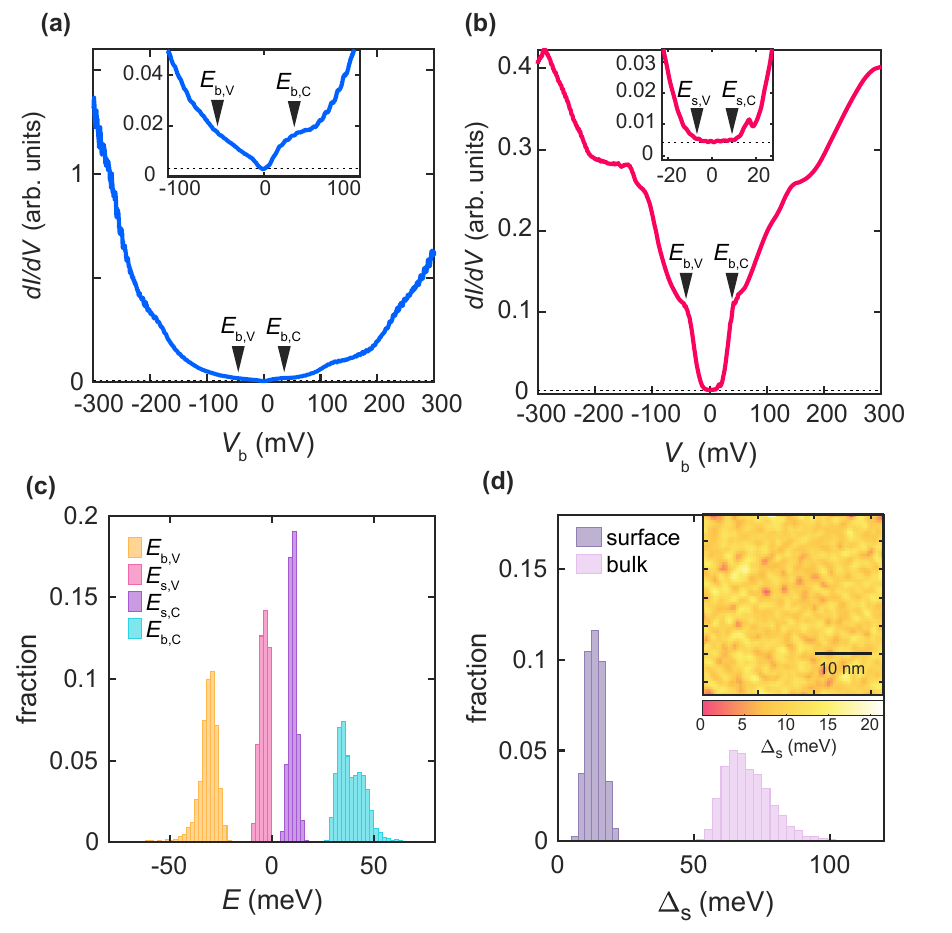}
\caption{\label{4Ksts} Spatial distribution of $\Delta_\mathrm{s}(\vec{r})$ at 4.2 K. (a) Representative STS measured at 4.2 K in which the tip's position was fixed with the bias voltage set at $V_\mathrm{b} = 100$ mV and tunneling current set to (a) $I_\mathrm{t} = 1000$ pA, and (b) $I_\mathrm{t} = 200$ pA. The inset reveals a zoomed in spectra of the low energy region. The horizontal dotted line is the baseline which indicates the minima of the spectra. (c) A histogram depicting the distribution of valence band maximum and conduction band minimum of the surface and bulk bands. (d) A histogram representing the distribution of $\Delta_\mathrm{s}(\vec{r})$ and $\Delta_\mathrm{b}(\vec{r})$. The inset exhibits the spatial map of $\Delta_\mathrm{s}(\vec{r})$.}

\end{figure}
Figures \ref{4Ksts}(a,b) exhibit representative $dI/dV$ spectra within experimental resolution taken at a single point on an atomically flat surface at 4.2 K with two different tunneling resistances. As $R_{t}$ is ramped from 100 M$\Omega$ (Fig. \ref{4Ksts}(a)) to 500 M$\Omega$ (Fig. \ref{4Ksts}(b)), the $dI/dV$ spectrum evolves from displaying dominantly bulk states, to mainly the surface states, with traces of bulk-derived features at higher energies. In contrast to a linear dependence of the LDOS as seen in the spectrum taken at 77 K (Fig. \ref{STS77K}(a)), the low energy region of the spectrum in Fig. \ref{4Ksts}(b)(inset) reveals a small energy range with zero LDOS. Such a zero-conductance energy region in the surface bands is a characteristic of a gap formation. The appearance of the gap at 4.2 K which was absent above $T_\mathrm{C}$, advocates that the gap formation associated with the topologically non-trivial surface states is a consequence of global ferromagnetic order perpendicular to the $ab$-plane \cite{PhysRevLett.111.146802}.

To assess the spatial variations of the topological surface gap, a $dI/dV$ scan over a (30 nm $\times$ 30 nm) area was measured with tunneling parameters, $I_\mathrm{t} = 200$ pA and $V_\mathrm{b} = 100$ mV. All $dI/dV$ spectra at this tunneling resistance revealed features from both the bulk and surface states consistently throughout the area as represented in Fig. \ref{4Ksts}(b). The average bulk band gap measured in this area is $\braket{\Delta_\mathrm{b}} (\vec{r})=(75.2 \pm 8.94)$ meV, 
with $\braket{E_\mathrm{b,V} (\vec{r})}=(-39.3 \pm 28.6)$ meV and $\braket{E_\mathrm{b,C}(\vec{r})}=(45.8 \pm 31.7)$ meV (Fig. \ref{4Ksts}(c,d)). The results obtained in this area at 4.2 K are consistent with those obtained in a different area at 77 K, except that the magnitude of the bulk gap at 4.2 K are marginally greater than that at 77 K, which is is to be expected, due to reduced thermal smearing at lower temperatures.

The surface gap is quantified through placing a baseline at each spectrum. The two energies at which the lock-in amplifier current begins to rise above the background dark current of the STM ($I_\mathrm{dark} = 1$ pA) from the baseline establish the upper and lower bounds of the surface gap.  Such careful analysis reveals that the average surface band gap on this area is  $\Delta_\mathrm{s}(\vec{r})=13.5$ meV and the region is gapped throughout the area with no gap-less sections (Fig. \ref{4Ksts}(c,d)). The highly homogeneous long-range ferromagnetic order in this scanned area of the MBE-grown MTI thin film investigated here is therefore in stark contrast to the findings reported in Ref. \cite{doi:10.1126/sciadv.1500740}, where a weakly interacting superparamagnetic domains of nanometer scale was observed. The surface gap values derived in this work are consistent with those reported in Refs. \cite{doi:10.1021/acsnano.6b03537, PhysRevB.101.245431}. 

\begin{figure*}[t]
\includegraphics[width =\textwidth]{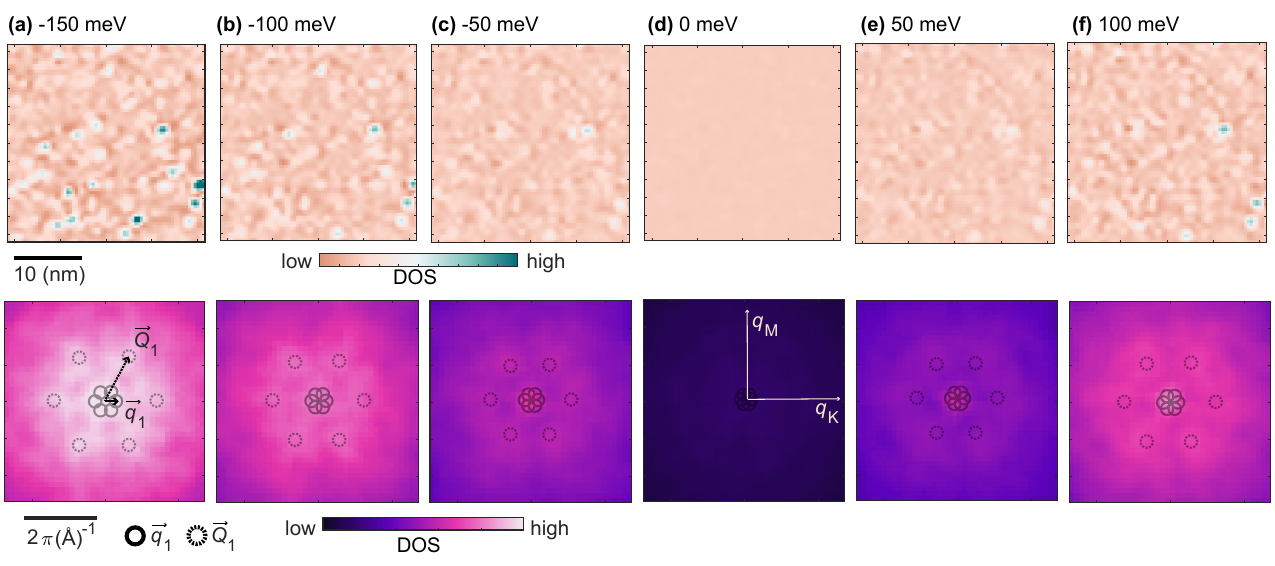}
\caption{\label{QPI} $dI/dV$ map at $T$ = 4.2 K. The top row panels exhibit the constant bias conductance map and the bottom row indicate the Fourier transformed images (QPI maps) for (a) $E=-150$ meV, (b) $E=-100$ meV, (c) $E=-50$ meV, (d) $E=0$ meV, (e) $E=50$ meV, and (f) $E=100$ meV. The colorbars indicate the LDOS intensity.}
\end{figure*}

\begin{figure*}[t]
\includegraphics[width =\textwidth]{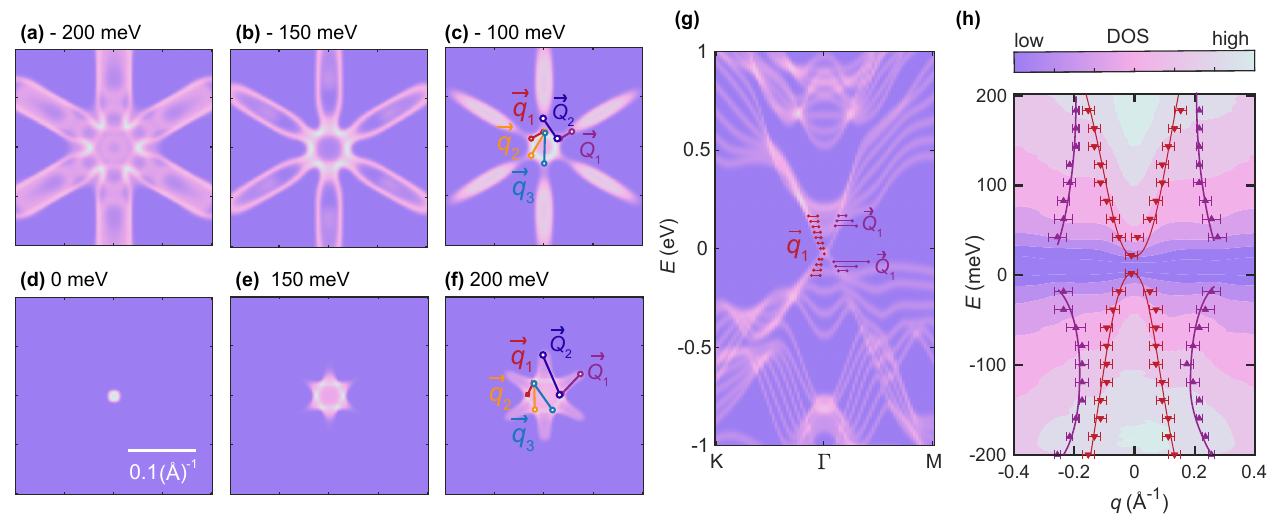}
\caption{\label{kDOS} Comparison of the calculated and measured energy dependence of LDOS in momentum space. Calculated momentum-resolved LDOS at (a) $E = -200$ meV, (b) $E = -150$ meV, (c) $E = -100$ meV, (d) $E = 0$ meV, (e) $E = 150$ meV and (f) $E = 200$ meV,. The horizontal axis in the maps is directed towards the K-point and the vertical axis is directed toward the M-point. Furthermore, the solid lines indicate the scattering vectors between pairs of high DOS regions on the map. (g) Calculated energy dependence of momentum resolved LDOS. The red solid lines indicate the energy evolution of $\vec{q_1}$ and the purple solid lines indicate the evolution of $\vec{Q_1}$. (h) A contour plot of the energy dependence of the measured QPI images taken along the M-$\Gamma$-M plane. The (red) $\triangledown$ symbols indicate the locations of the  $\vec{q_1}$ scattering wavevectors, and the (purple) $\triangle$ symbols indicate the location of the $\vec{Q_1}$ scattering wavevectors. }
\end{figure*}
To further support the distinction made between the surface and the bulk gaps, momentum dependent information is extracted from the $dI/dV$ spatial maps by taking the tunneling conductance at a constant bias voltage for each pixel and then mapping out the conductance over a two-dimensional (2D) area. The top panels of Fig. \ref{QPI}  exemplify a constant bias LDOS maps at selected energies of $E = -150$ meV, $-100$ meV, $ -50$ meV, $0$ meV, $50$ meV and $100$ meV. Noting that the local fluctuations in the LDOS at a constant energy represented standing waves of quasiparticle interferences induced by impurity scatterings from dopants and defects, 2D Fourier transformation of the LDOS spatial maps at constant energies yielded information about the impurity scattering wavevectors of quasiparticle interference (QPI) images in momentum space. The QPI images associated with the constant energy LDOS maps are shown in the bottom panels of Fig. \ref{QPI}, each exhibiting a single bright spot in the $\Gamma$-point and six spots around it in the M-direction.

For a full understanding of the correspondence between the scattering wavevectors and the patterns the QPI images, all possible scattering vectors on the Fermi surface, not only the intra-band surface state scattering wavevectors ($\vec{q}$) but also the inter-band scattering wavevectors ($\vec{Q}$) between the surface and the bulk states must be considered, given that spectral features associated with both the bulk and surface states were present in the $dI/dV$ spectra (Fig. \ref{4Ksts}). In general, the prominent scattering vectors observed in QPI images are those scattering between pairs of momentum regions on the Fermi surface with high LDOS \cite{Beidenkopf2011, PhysRevB.80.245439, PhysRevLett.106.206805, PhysRevLett.103.266803}.

Figure \ref{kDOS}(a-g) indicates the theoretical momentum ($\vec{k}$)-dependent constant energy LDOS maps of 6-QL Sb$_2$Te$_3$, which is a good approximation to 6-QL (Bi$_{0.2}$Sb$_{0.8}$)$_2$Te$_3$, obtained through density functional theory (DFT) calculations. There are three possible intra-band scattering wavevectors amongst the high LDOS regions on the surface band, which are denoted as  $\vec{q_1}$,  $\vec{q_2}$, and  $\vec{q_3}$ , as shown in Figs. \ref{kDOS} (c,f). 

The scattering across the opposite nodes on the Dirac cone,  $\vec{q_3}$, is prohibited by Kramer’s rule, and so can be excluded. Moreover, a previous STS study presented in Ref. \cite{PhysRevB.80.245439} has indicated the absence of $\vec{q_2}$ and found $\vec{q_1}$ to be the prominent dominant intra-band scattering in the surface bands. The diameter of the Dirac cone at 100 meV below $E_\mathrm{D}$ is approximately $0.05\mathrm{\AA} ^{-1}$ for Sb$_2$Te$_3$ as indicated by the DFT calculations (Fig. \ref{kDOS}(c)) as well as the angle resolved photoemission spectrum (ARPES) measurements \cite{Zhu2015, PhysRevB.90.081106}. The scattering wavevector from the adjacent nodes of the hexagonal Dirac cone of the surface state at $E = -100$ meV is therefore approximately $\vec{q_1}=0.03$ $\mathrm{\AA}^{-1}$ from Fig. \ref{kDOS}(c). Hence, considering the scale of the scattering wavevectors, we associate the spots clustered around the $\Gamma$-point along the $\Gamma$-M direction in the QPI image with the $\vec{q_1}$ scattering wavevectors (Fig. \ref{QPI} (c)).

By stacking up $dI/dV$ maps of each bias voltages and taking a slice across the $E-q$ plane, an energy dependent contour map is obtained (Fig \ref{kDOS}(h)). The energy dependence of $|\vec{q_1}|$ can be derived from this $E-q$ contour plot across the M-$\Gamma$-M direction. Evidently, $|\vec{q_1}|$ in the M direction contracts as the energy evolves from below $E_\mathrm{F}$ and then expands as the energy increases above $E_\mathrm{F}$, which is in accordance with the energy dependence of the radius of the surface-state Dirac cone (Fig. \ref{kDOS}(g)), thus, further verifying the identification of the $\vec{q_1}$ scattering wavevector of the  QPI maps. 

Although the QPI maps also show clear defect-mediated surface-to-bulk scattering at  energies of 100 meV above and below $E_\mathrm{F}$, the absence of $\vec{Q_1}$ close to $E_\mathrm{F}$ asserts that quasiparticles are largely confined to within the surface near $E_\mathrm{F}$, and bulk states are well separated from the Fermi level. We can consequently rule out the possibility of significant bulk states contributing to the impairment of quantized transport.

\subsection{Discussions}

\subsubsection{The role of defects in spatial modulations of $\Delta_s$}

\begin{figure}[b]
\includegraphics[width =0.48\textwidth]{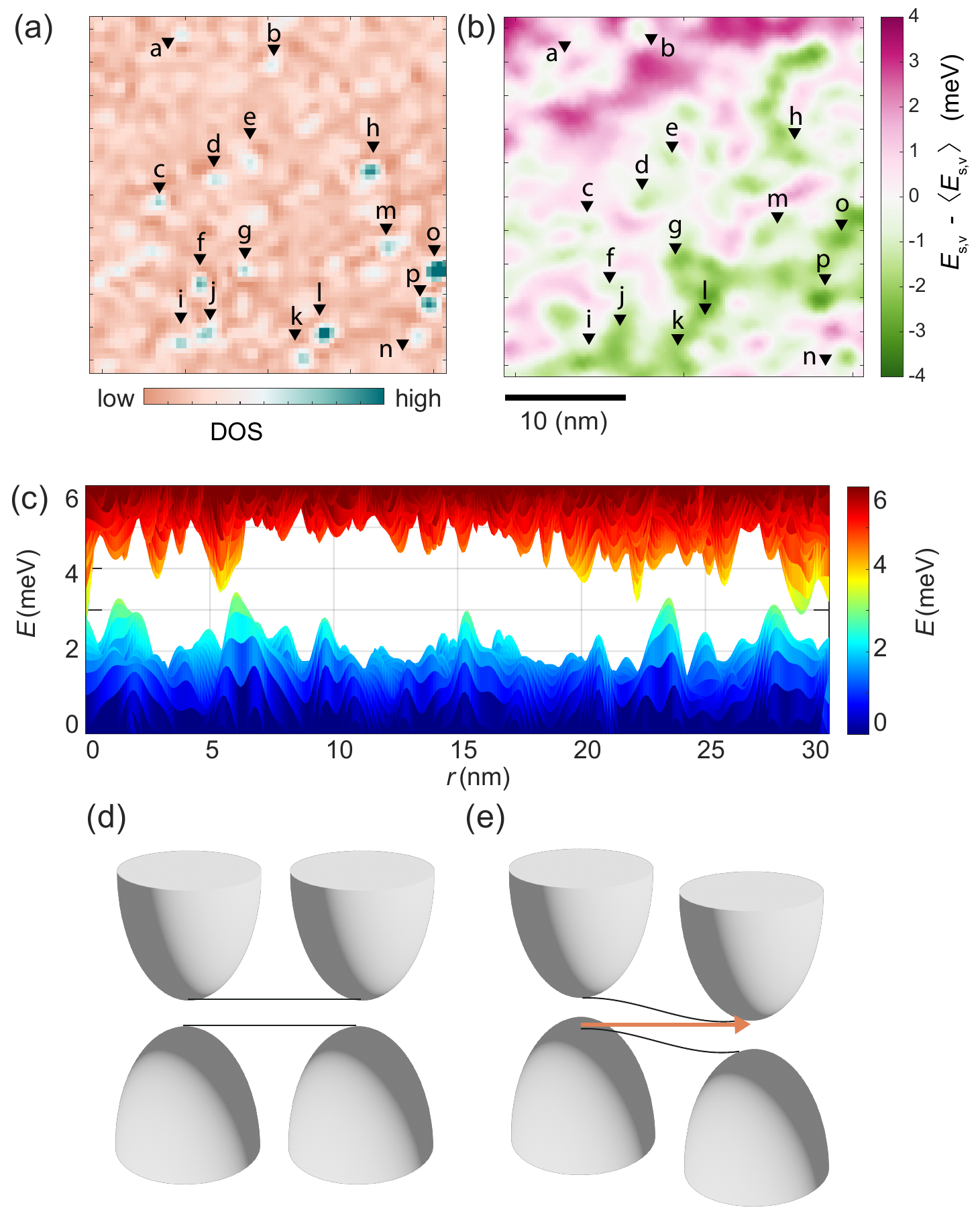}
\caption{\label{gapmodulation} Spatial profile of the topological surface gap. (a) A constant bias conductance map corresponding to an energy of $E = -150$ meV. The arrows point to representative defect locations found in this area. (b) A spatial distribution map of $E_\mathrm{s,V}$ with respect to its average, $\braket{E_\mathrm{s,V}}$ in the same area of panel (a).  The arrows indicate local minima in the $\braket{E_\mathrm{s,V}}$ and the approximate defect location of the defects labelled in panel (a). (c) Projection of a surface plot depicting the spatial profile of the topography and (b) its corresponding spatial profile of $E_\mathrm{s,V}$ and $E_\mathrm{s,C}$ obtained from STS scans over a 30 nm region at $T = 4.2$ K. Although the average surface gap at each point in the area is 13.5 meV, the spatial profile exhibits an effectively smaller gap owing to the energy modulations. Schematic diagram illustrating (d) an ideal case in which the surface gap through space are all aligned, and for (e) the case in which spatial modulations in $E_\mathrm{s,V}$ and $E_\mathrm{s,C}$ can allow in-direct gap hopping, yielding an effectively gapless metallic phase on a nanoscale range at $T = 4.2$ K.}
\end{figure}

Both point spectroscopy and QPI analyses suggest that the bulk bands are well separated from the surface bands, granting direct thermal excitation of electrons either from bulk to bulk or from bulk to surface states on a local scale energetically unfavorable at least at 4.2 K. Rather than being inherent in the magnitude of $\Delta_s$, spatial map of $\Delta_s$ indicates that the problem of vanishing QAHE at slightly elevated temperature lies in its spatial alignment of energy bands. Given that the average surface gap of the sample is approximately 13.5 meV, local modulations in the sample due to inhomogeneous doping, defects, or impurities of a few meVs lead to misalignment of the surface energy bands, allowing for tunneling of electrons from the surface valence band of a particular point to the surface conduction band of another separate point. Indeed, it is found that the energy of the surface conduction band in one region can be lower than the surface valence band in another region as shown in Fig. \ref{gapmodulation}(c). 

To discern the origin of the observed energy modulation, the spatial distribution of $E_\mathrm{s,V}$  was compared with the positions of defects, revealed from the constant bias conductance map presented in Fig. \ref{QPI}(a). Given that quasiparticles experience scattering off defects, these defects induce modulations in the conductance map, facilitating their detection, as depicted in Fig. \ref{gapmodulation}(a). By cross-referencing the locations of defects with the distribution of $E_\mathrm{s,V}$ measured from the same area (Fig. \ref{gapmodulation}(b)), a clear correlation emerges: the positions of defects align with distinct energy levels of $E_\mathrm{s,V}$. Notably, the regions populated with defects coincide with lower energy levels, as indicated by the green patches in Fig. \ref{gapmodulation}(b).

This correlation strongly suggests that the observed modulations in local chemical potentials are a direct consequence of the inherent defects within the system, yeilding a spatially modulated gap, as outlined in the schematic diagram of Fig. \ref{gapmodulation}(d,e).

As means of verifying the presence of a leaky current running along the surface state due to the energy modulation of bands, the tunneling probability of electrons across the topological surface as a function of disorder is numerically investigated in a finite super-lattice Hamiltonian of pristine Sb$_2$Te$_3$. 
Although there are various forms of defects in the sample including Sb/Te vacancies, antisites, typically in MBE grown samples, the concentration of such is $10^{-4}$ nm$^2$ \cite{PhysRevLett.108.066809}, which is significantly dilute than the concentration of dopant atoms. For this reason, the main source of disorder within a nanometer scale region come from Cr- and Bi-dopants, which have vastly different atomic mass and orbital characters from atoms in Sb$_2$Te$_3$ compound \cite{PhysRevB.92.195418}. The effect of disorder generated by the dopant atoms are incorporated by emulating the shift in the local chemical potential ($\delta \mu^\alpha$, ($\alpha=$Cr,Bi)) introduced through Sb-site substitution. Since each Sb-site orbital has six orbitals (three $p$-orbitals for each spin configuration), each dopant shifts the chemical potential for all six orbitals at each Sb-site.

Assuming an RKKY interaction, in this model each orbital of the Cr-dopant contributes to a $z$-component in flux density with a $B_z \propto r^{-3}$ dependence from its Wannier center in the radial direction \cite{PhysRevLett.102.156603}. Given that the Wannier orbitals of Cr-dopants substitute a total of $N_\mathrm{Cr}$ number of Sb-site orbitals located at $\vec{R_i}$, each Cr-dopant orbital can be approximated to collectively contribute to a superposition of flux density,

\begin{figure}[b]
\includegraphics[width =0.5\textwidth]{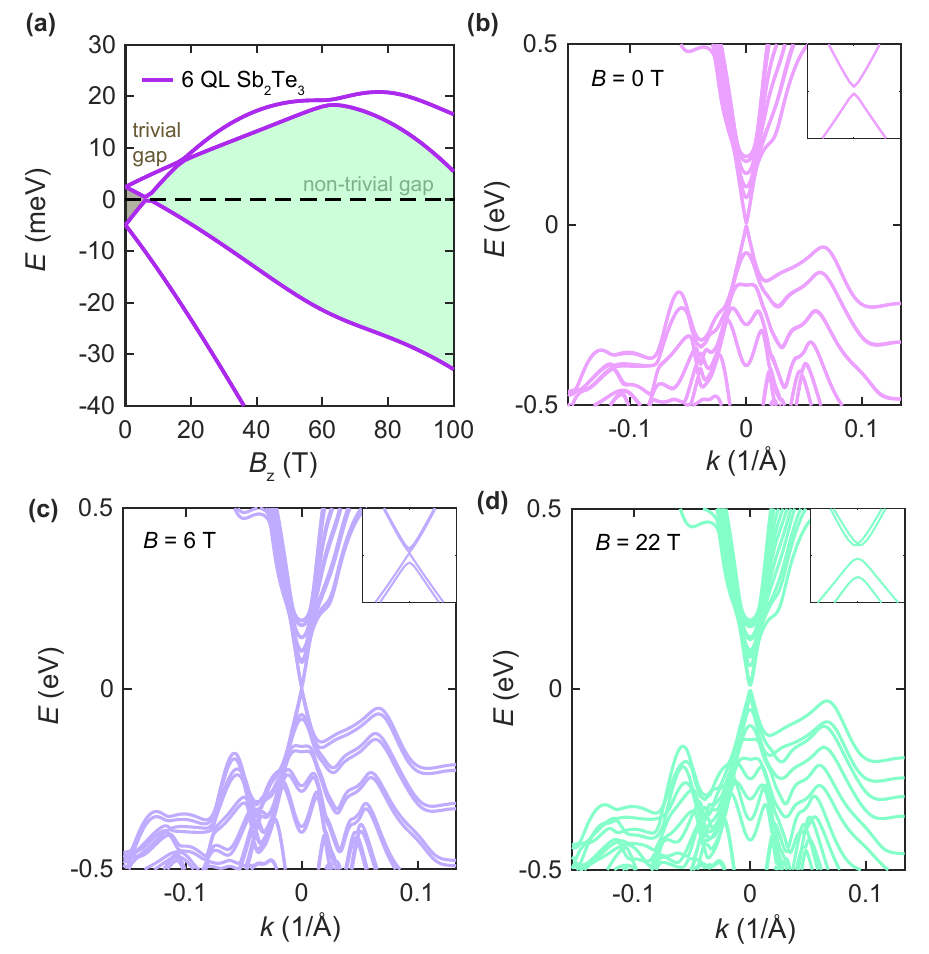}
\caption{\label{bandevolution} Surface gap dependence of magnetic field in 6 QL Sb$_2$Te$_3$. (a) The evolution of the lowest two conduction bands and the highest two valence bands as a function of magnetic field. The corresponding band structure of 6 QL Sb$_2$Te$_3$ along the K-$\Gamma$-M path at (b) $B = 0$ T, (c) $B = 6$ T, where the gap closes, and (d) $B = 22$ T where a topologically non-trivial magnetic gap is observed.  The inset exhibits a blow-up of the low energy region, in which the vertical axis scales from $-50$ to $50$ meV and the horizontal axis scales from $- 0.01$ $\mathrm{\AA}^{-1}$ to $0.01$ $\mathrm{\AA}^{-1}$. }
\end{figure}

\begin{align} \label{Bz}
B_z(\vec{r}) &= B_{z0} \sum_i^{N_\mathrm{Cr}} \bigg(\frac{1-e^{-|(\vec{r}-\vec{R_i})/r_0|^3}}{|(\vec{r}-\vec{R_i})/r_0|^3} \bigg) \hat{r},
\end{align}
where $B_{z0}$ is the maximum flux density induced by a single local spin polarization and $r_0$ is the range of the magnetic exchange interaction, which is set to be $r_0 = 15$ nm\cite{PhysRevLett.102.156603}. A single dopant therefore creates a nearly uniform field, given that $r_0$ is  substantially longer than the lattice parameter, spanning over 30 times the width of a unit cell. The magnetic field coming from Cr-dopants has two effects: (1) induce spin-energy splitting through Zeeman interaction, as well as (2) altering the electron motion through a Peierl's substitution.  Not only contributing to magnetization, Cr-dopants also have an effect of altering the Fermi level with respect to the Dirac point, which is reflected in the theoretical model as a shift in the onsite energy ($\varepsilon_{ii}$).  For  Cr-dopants, this shift is approximately $ \delta \mu^\mathrm{Cr} (\vec{r}) = +150$ meV, as revealed through ARPES and STS studies \cite{PhysRevB.92.195418, PhysRevB.98.115165}. Furthermore, to enforce a ferromagnetic insulating state \cite{doi:10.1126/science.1187485},  hopping to Cr-dopants are prohibited in the model. All in all, the change in Hamiltonian owing to magnetic dopants is
\begin{align} \label{HCr}
\hat{H}^\mathrm{Cr}(\vec{B}) =  \sum_i^{N} g_{||} \mu_B B_z(\vec{r}) \psi_i^\dagger \sigma_z \psi_i + \sum_{i'}^{N_\mathrm{Cr}}\delta \mu ^\mathrm{Cr} (\vec{r})  \psi_{i'}^\dagger \psi_{i'} \notag \\
+  \sum_{i'}^{N_\mathrm{Cr}}\sum_{i\neq j}^{N} t_{ij}\cdot e^{i\Theta_{ij} } \bigg( \psi_i^\dagger \psi_j + \psi_j^\dagger \psi_i \bigg) (1-\delta_{ii'})
\end{align}
where $g_{||}$ is the Land\`e $g$-factor parallel to the crystal $c$-axis, which is approximately 25 generally for 3D topological insulators \cite{PhysRevB.93.155114, Sakhin2021}, $t_{ij}$ is the hopping energy and $\Theta_{ij}$ is the Peierl's phase, given by
\begin{align}
\Theta_{ij} & = \frac{h}{e}\int_{\vec{R_i}}^{\vec{R_j}} \frac{\vec{r} \times \vec{B}}{2} \cdot{d\vec{r}}.
\end{align}
\begin{figure}[t]
\includegraphics[width =0.5\textwidth]{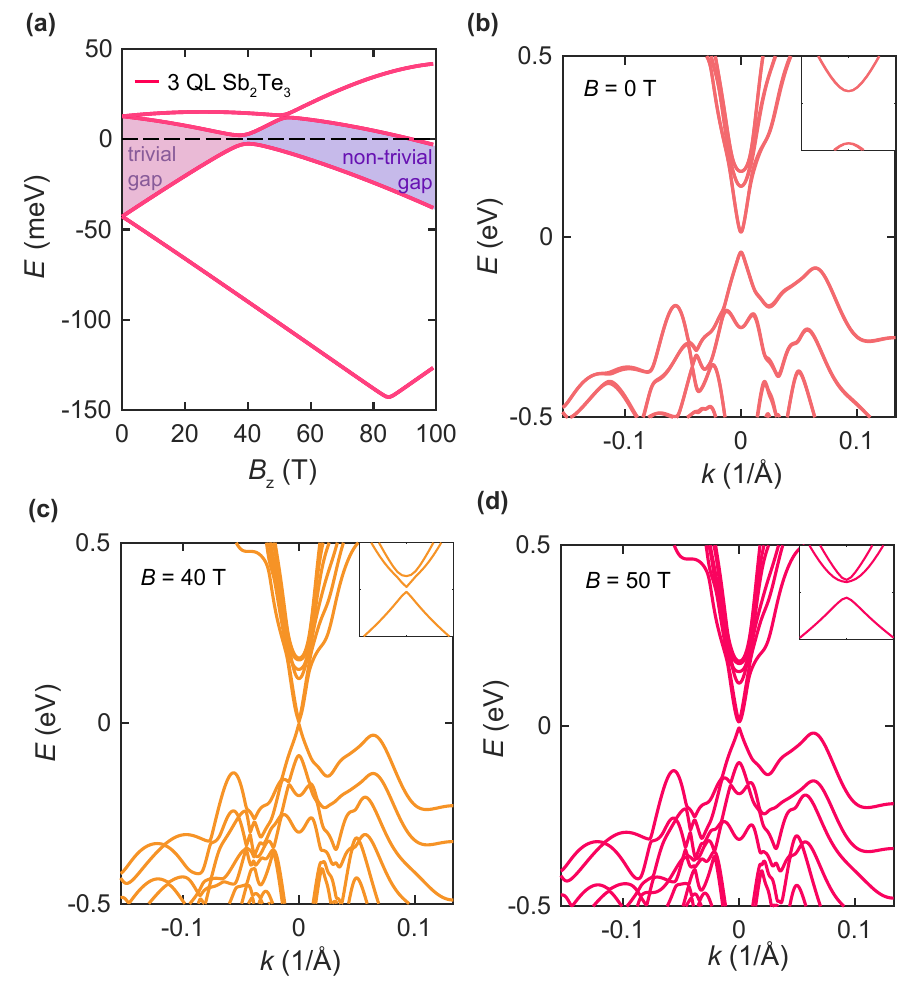}
\caption{\label{bandevolution3QL} Surface gap dependence of magnetic field in 3 QL Sb$_2$Te$_3$. (a) The evolution of the lowest two conduction bands and the highest two valence bands as a function of magnetic field. The corresponding band structure of 3 QL Sb$_2$Te$_3$ along the K-$\Gamma$-M path at (b) $B = 0$ T, (c) $B = 40$ T, where the gap closes, and (d) $B = 50$ T where a topologically non-trivial magnetic gap is observed.  The inset exhibits a blow-up of the low energy region, in which the vertical axis scales from $-50$ to $50$ meV and the horizontal axis scales from $- 0.01$ $\mathrm{\AA}^{-1}$ to $0.01$ $\mathrm{\AA}^{-1}$. }
\end{figure}
For the case of Bi-dopants, which do not contribute to magnetization, only the on-site energies are altered and the hopping terms are left untouched
\begin{align} \label{HBi}
\hat{H}^\mathrm{Bi}(\vec{r}) =  \sum_{i'}^{N_\mathrm{Bi}}\delta \mu ^\mathrm{Bi} (\vec{r})  \psi_{i'}^\dagger \psi_{i'} \notag.
\end{align}
Here $N_\mathrm{Bi}$ represents the number of Wannier orbitals of the Sb-site subsituted with the Wannier orbitals of the Bi-dopants, and the shift in the onsite energy is $\delta \mu^\mathrm{Bi} = - 100$ meV, determined through ARPES measurements \cite{Zhang2011}. 

Fig. \ref{bandevolution} exhibits the bandstructures of Sb$_2$Te$_3$ in the limit where there is uniform magnetic field threaded through the lattice and in the case in which there are no disorder. As the magnetic field increases, degeneracy of the bands are lifted and a band inversion occurs at around a flux density of $B = 6$ T (Fig. \ref{bandevolution}(a,c)), thus giving rise to a magnetic gap at higher fields. This critical field at which a band inversion occurs is consistent to the order of magnitude of the critical field $B_\mathrm{C} \approx 20 \mathrm{meV}/(25\mu_B) \approx 10$ T computed by a 4-band model of 5QL Bi$_2$Se$_3$ topological insulator thin film \cite{doi:10.1126/science.1187485}. With a field reaching $B \approx 22$ T, the magnetic gap is $\approx 15$ meV, which is approximately about the magnitude of $\Delta_\mathrm{s}$ experimentally seen at 4.2 K. Notably, even with extreme magnetic fields of 100 T, the magnetic gap only reaches as high as $\approx 40$ meV.

Likewise, 3 QL Sb$_2$Te$_3$ also shows a band inversion but at a higher field of $B = 40 $ T owing to larger hybridization gap that necessitates a larger Zeeman splitting (Fig. \ref{bandevolution3QL} ). The size of the surface gap is smaller in general compared to the 6 QL case. At a higher field of 50 T, the gap reaches a magnitude $\approx 10 $ meV, however, the surface bands are less separated from the bulk bands (Fig. \ref{bandevolution3QL}(d)). Thus, incorporating a field within a window larger than $B = 40$ T, but less than $B = 50$ T ensures a QAH-phase in the 3 QL system in which the surface is well separated from the bulk states. Given the correspondence between the 3 QL and the 6 QL systems, the effects of disorder and phonon-coupling can therefore be estimated in the QAHE-phase of a 3 QL system and then compared to experimental results of 6 QL samples to maintain reasonable computational costs.

\begin{figure}[t]
\includegraphics[width =0.5\textwidth]{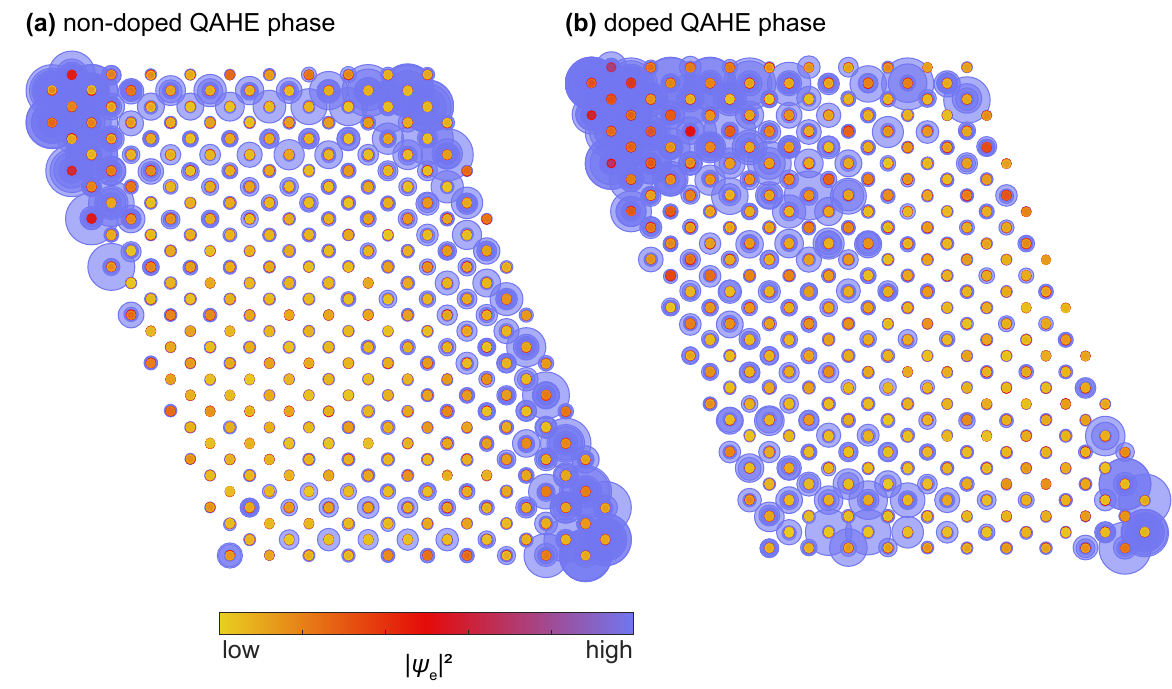}
\caption{\label{bubble} A bubble chart depicting the edge state probability density $|\psi_\mathrm{e}|^2$ in a ($10\times10\times1$) 3 QL finite lattice at their corresponding Wannier centers projected along the crystal $ab$-plane (a) without any defects ($i.e.$, $N_\mathrm{Bi} = N_\mathrm{Cr} = 0$ but in the QAH-state by turning on magnetization with $B_\mathrm{z} = 45$ T), and (b) 108 Bi-dopants and 60 Cr-dopants  ($i.e.$, $N_\mathrm{Bi} = 648$ and $N_\mathrm{Cr} = 360$ with $\braket{B_z} \approx 45$ T).}
\end{figure}

Effects of disorder were implemented to a finite ($10\times10\times1$) superlattice 3 QL system consisting of 1500 atoms ($N = 9000$ total orbitals) and randomly replacing the orbitals of Sb-sites with 108 Bi-dopants ($x = 0.8$, $N_\mathrm{Bi} = 648$) and 60 Cr-dopants ($y = 0.1$, $N_\mathrm{Cr} = 360$), where each Cr orbital contributes to a flux density of $B_0 = 0.125$ T (such that the average total flux density $\braket{B_z} \approx 45$ T).
To assess the effects of modulations in local chemical potential owing to the dopants to the chiral edge states, we compare the spatial distribution of the surface state probability density in the finite lattice of two cases, (a) \textit{non-doped QAH-phase}: no dopants ($N_\mathrm{Bi} = N_\mathrm{Cr} = 0$, yet in the QAH-state by turning on magnetization with ${B_z} = 45$ T), and (b) \textit{doped QAH-phase} the case with both Bi- and Cr- dopants matching the stochiometry of \sample.

Specifically, this is visualized by plotting the probability density of all the edge eigenstates ($|\psi_\mathrm{e}|^2$) within an energy window of $\pm 5$ meV from $E_\mathrm{F}$,
\begin{align}
\ket{\psi_{\mathrm{e}}(\vec{r_i})} =  \sum_{\substack{|E_n-E_{\hspace{1 pt}\mathrm{F}}| \\ <  5 \mathrm{meV}}} \ket{\psi_n(\vec{r_i})}.
\end{align}
In the lattice without any dopants, states within this 10 meV energy frame where the topological gap is located is predominantly clustered near the edge and virtually no states towards the center of the crystal (Fig. \ref{bubble}). As defects are incorporated (Fig. \ref{bubble}(b)), patches of substantial $|\psi_\mathrm{e}|^2$ are observed in the interior of the crystal. Therefore, dopants play a role in modulating energies of eigenstates close to $E_\mathrm{F}$.

\subsubsection{The role of phonon mediated band renormalization}

\begin{figure}[t]
\includegraphics[width =0.5\textwidth]{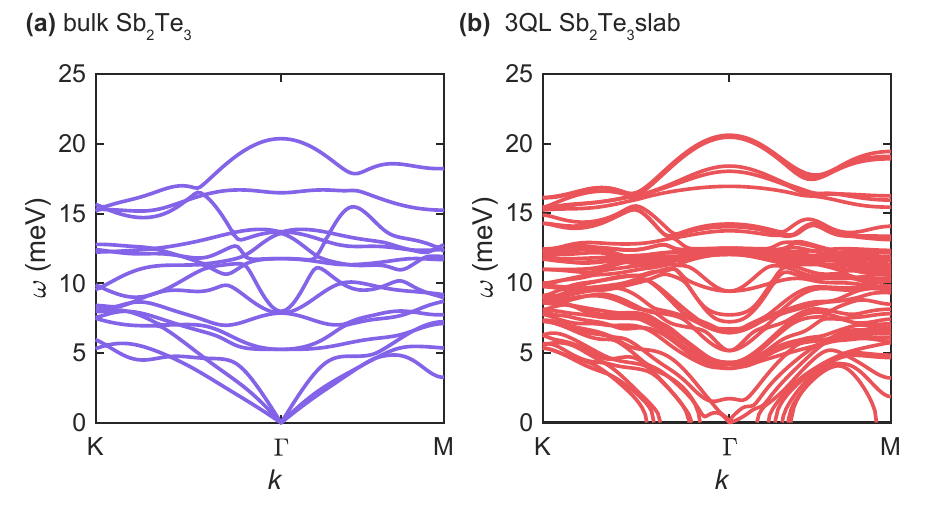}
\caption{\label{phonon} Phonon spectrum along the K-$\Gamma$-M path of Sb$_2$Te$_3$ for (a) a primitive cell and (b) a 3 QL slab geometry.}
\end{figure}

The theory of topological insulators have been built upon a non-interacting quasiparticle framework. However, when putting into consideration effects of finite temperature, which cannot be neglected in experiments, there is no escape from recognizing many-body interactions \cite{Heid2017}. In particular, at finite temperatures, electron-phonon ($e-p$) interactions are dominant scattering mechanism of the surface state Dirac electrons \cite{PhysRevLett.108.185501}. Inclusion of an $e-p$ term in the Hamiltonian of the system can consequently renormalize the electronic band structure in topological insulators \cite{PhysRevLett.110.046402}. Although there is no consensus on the $e-p$ coupling strength between phonons and the topological surface states due to its experimental difficulties \cite{Heid2017, PhysRevLett.108.185501, PhysRevLett.110.046402, PhysRevB.83.241303, PhysRevLett.108.187001, PhysRevLett.110.217601, PhysRevB.85.035441, Benedek2020, https://doi.org/10.1002/pssb.2220840226, PhysRevLett.113.157401}, experimental methods including ARPES have directly observed phonon scattering in the surface states \cite{Heid2017, PhysRevLett.108.185501, https://doi.org/10.1002/pssb.2220840226, PhysRevLett.113.157401}.

Figure \ref{phonon} exhibits the phonon spectrum of the bulk and slab geometry of Sb$_2$Te$_3$. Since there are 5 atoms in a primitive unit cell, there are 15 phonon modes, in which 3 are low energy acoustic modes, and 12 are higher energy optical modes. To study the $e-p$ interactions amongst the surface states, phonons were calculated for a 3 QL slab geometry (Fig. \ref{phonon}(b)). In a 3 QL slab, there are a total of 15 atoms, resulting in a sum of 45 phonon modes. There are numerous acoustic modes that extend below 0 frequency, which reflect imaginary phonon modes emerging from anharmonicity. Usually, the existence of imaginary modes is associated with dynamical instabilities such as the Jahn-Teller instability. However, the rise of imaginary phonon modes in this specific case can be considered a result of phonon anharmonicity with the presence of a vacuum above and below the slab geometry in which the phonon modes were calculated.  

The electron-phonon interaction to the first order is expressed as 

\begin{align} \label{Hep}
\hat{H}^{ep} (\vec{r}) &= \frac{1}{M\sqrt{N}} \sum_{\vec{k} \vec{q}} \sum_{m n \nu}g_{m n \nu}(\vec{k},\vec{q})c^\dagger_{m,\vec{k}+\vec{q}} c_{n,\vec{k}} x_{q\nu}, 
\end{align}
where $x_{q,\nu} = a_{\nu,-\vec{q}} + a^\dagger_{\nu,-\vec{q}}$ is the bosonic displacement operator \cite{RevModPhys.89.015003}. Moreover, $M$ is the total number of bands, $N$ is the number of unit cells, $c^\dagger$  and $c$ ($a^\dagger$  and $a$) are the associated fermionic (bosonic) creation and annihilation operators for electrons (phonons), respectively . Notably, this is a single-phonon absorption/emission process in which momentum is conserved and the amount of energy transfer between the electron and phonon is denoted by the coupling matrix $g_{mn\nu}$. 

\begin{figure}[t]
\includegraphics[width =0.5\textwidth]{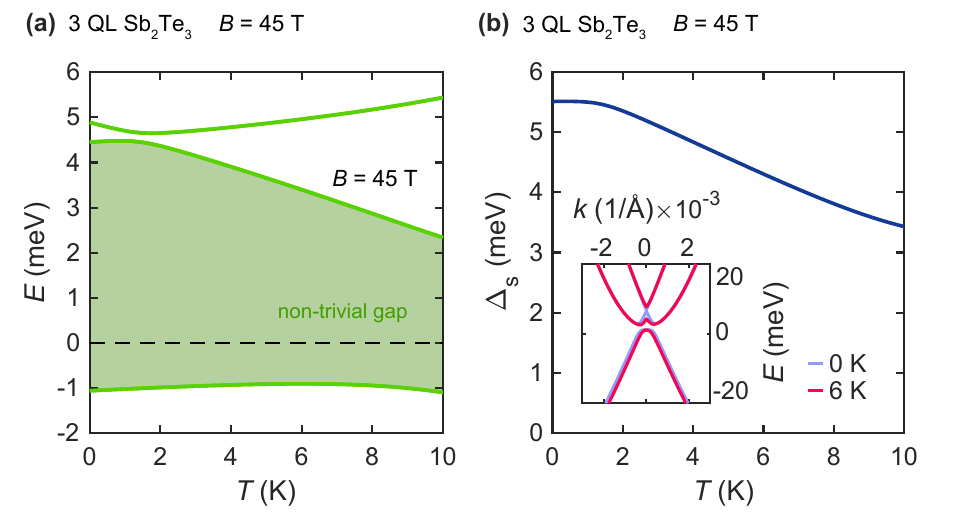}
\caption{\label{tempdependence} Effects of $e-p$ renormalization on the surface state band structure. (a) The temperature dependence of surface valence and conduction bands of 3 QL Sb$_2$Te$_3$ with a magnetic field of 45 T to create a QAH-phase. (b) The temperature dependence of $\Delta_\mathrm{s}$. The inset shows the surface state band structure around the $\Gamma$-point for $T = 0$ K and $T = 6$ K. }
\end{figure}

Through a change of basis by Fourier transforming creation/annihilation operators into field operators,
\begin{align}
c^\dagger_{m,\vec{k'}} &= \frac{1}{\sqrt{N}}\sum_{i} \psi_m^\dagger(\vec{r_i})e^{+i(\vec{k} + \vec{q})\cdot \vec{r_i}}  \\
c_{n,\vec{k}} &= \frac{1}{\sqrt{N}}\sum_{j} \psi_n(\vec{r})e^{-i\vec{k}\cdot \vec{r_j}},
\end{align}
the real-space Hamiltonian is represented as
\begin{align}
\hat{H}^\mathrm{ep}(\vec{r}) & =  \frac{1}{M N\sqrt{N}} \sum_{i,j} \sum_{mn\nu} g_{mn\nu}(\vec{r_i},\vec{r_j})  \psi_m^\dagger(\vec{r_i}) \psi_n(\vec{r_j})
\end{align}
where the electron-phonon coupling matrix element in the lattice representation is
\begin{align} \label{BE}
g_{mn\nu}(\vec{r_i},\vec{r_j}) & = \sum_{\vec{k} \vec{q}} g_{mn\nu}e^{i\vec{k}\cdot(\vec{r_i}-\vec{r_j})+i\vec{q}\cdot\vec{r_i}} \Delta x_{q \nu}.
\end{align}
Given that the linear component of electron-phonon interactions is irrelevant to band renormalization \cite{P_B_Allen_1976}, we consider that in thermal equilibrium the phonon interaction rather coming from phonon displacement $x_{q\nu}$, emanates from the variance $\Delta x_{q\nu} = \sqrt{(\braket{x^2} - \braket{x}^2} = \sqrt{ \hbar \omega_\mathrm{ph} n_\mathrm{BE}}$, thereby accounting for fluctuations in atomic displacement. Here, $n_\mathrm{BE} = 1/ (e^{\hbar \omega_\mathrm{ph}/k_\mathrm{B}T}+1)$ is the Bose-Einstein distribution function, expressing the phonon population. Using the $e-p$ coupling constant $g_{mn\nu}$ obtained through DPFT calculations, temperature dependent effects on the surface states were computed.

As the temperature of the system increases from 0 to 10 K, a recognizable reduction of meV-order in $\Delta_\mathrm{s}$ is observed (Fig. \ref{tempdependence}).
Within this temperature range, $\Delta_\mathrm{s}$ remains gapped, implying that at liquid-$^4$He temperature of 4.2 K, QAH-phase is present (Fig. \ref{tempdependence}). Nonetheless, such renormalization effect which induces a gap reduction augments vulnerablility of the system to rise of leaky currents due to disorder-induced spatial energy modulations of the eigenstates.

\subsubsection{Combined effects of electron-phonon interaction and disorder on surface band renormalization }

In experimental reality, finite temperature effects and disorder are simultaneously present. To take both into account, Eq. \eqref{Hep} and Eq. \eqref{HCr} are introduced as perturbations, such that the fully integrated Hamiltonian is
\begin{align}
\hat{H}(\vec{B}, T) & = \sum_{i,j}^{N} \varepsilon_{ii} \psi_i^\dagger \psi_i  + \hat{H}^\mathrm{Cr}(\vec{B}) + \hat{H}^\mathrm{Bi}(\vec{r}) + \hat{H}^\mathrm{ep}(T),
\end{align}
which is implemented to study the collective effects of disorder and $e-p$ interactions on a ($5\times 5\times1$) finite super-lattice of 3 QL Sb$_2$Te$_3$.

\begin{figure}[t]
\includegraphics[width =0.45\textwidth]{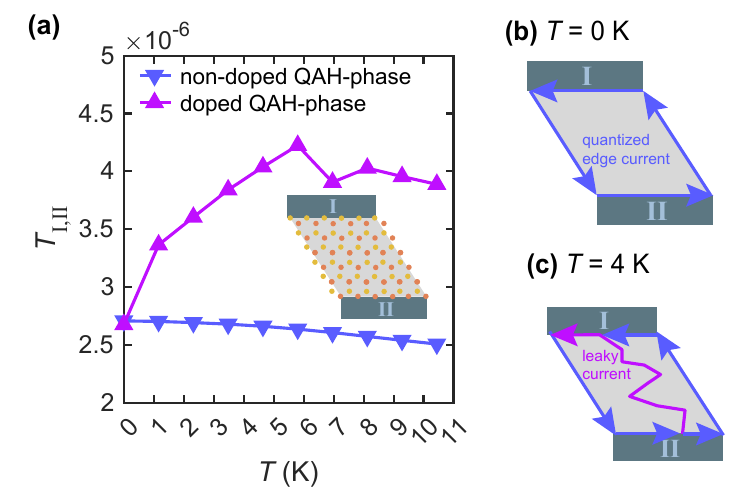}
\caption{\label{device} Effects of $e-p$ renormalization on the surface state band structure. (a) T$_\mathrm{I,II}$ as a function of temperature with no dopants yet in the QAHE state owing to application of magnetization of $B_\mathrm{z} = 45$ T and incorporation of 15 Bi-dopants and 27 Cr-dopants with $\braket{B_\mathrm{z}} = 45$ T to emulate the stochiometry of \sample. The inset illustrates a schematic diagram of the ($5\times5\times1$) super-lattice for calculating probability that an electron can be transmitted across the crystal from edge I to II. (b) A schematic diagram illustrating quantized transport at $T = 0$ K, and (c) compromised quantized transport at finite temperatures as a consequence of both defect and phonon mediated metal transition.}
\end{figure}

Whether electrons can tunnel through inhomogeneous puddles of states that form inside the topological surface gap owing to such many-body interaction is assessed through calculating the transfer matrix $T_\mathrm{I,II}$ for electrons from 
one edge (I) reaching to the opposite edge (II), as illustrated in Fig. \ref{device}(a). For both the non-doped QAH-phase and doped QAH-phase, the transfer matrix $T_\mathrm{I,II}$ is given by 
\begin{align}
 T_\mathrm{I,II} & = \mathrm{Tr}\big[ \hat{\rho} (t)\rho_\mathrm{II} \big]  = \mathrm{Tr}\big[ \hat{U} \hat{\rho}_\mathrm{I}\hat{U}^\dagger\hat{\rho}_\mathrm{II} \big] ,
\end{align}
Here the density operators, $\hat{\rho}_\mathrm{I}$ and $\rho_\mathrm{II}$, express the edge state of surface state orbitals on edge I and edge II,
\begin{align}
 \hat{\rho}_\mathrm{I} &= \frac{1}{N_\mathrm{I}}\sum_{\substack{|E_n-E_{\hspace{1 pt}\mathrm{F}}| \\ <  5 \mathrm{meV}}} \sum_\mathrm{I}\ket{\psi_n(\vec{r_\mathrm{I}})}\bra{\psi_n(\vec{r_\mathrm{I}})}  \\
 \hat{\rho}_\mathrm{II} & =  \frac{1}{N_\mathrm{II}}\sum_{\substack{|E_n-E_{\hspace{1 pt}\mathrm{F}}| \\ <  5 \mathrm{meV}}} \sum_\mathrm{II}\ket{\psi_n(\vec{r_\mathrm{II}})}\bra{\psi_n(\vec{r_\mathrm{II}})},
\end{align}
where $\hat{U}_n = e^{-iE_nt/\hbar}\ket{\psi_n}\bra{\psi_n}$ is the unitary time-evolution operator of the $n$-th eigenstate. The time scale $t$ is set to a value large enough to allow for $T_\mathrm{I,II}$ to saturate to its maximum for all temperature ranges.  
Figure \ref{device}(a) exhibits that by excluding dopants, the transmission probability across the device is virtually unchanged with increasing temperatures. In other words, the role of $e-p$ interactions by itself are not sufficient for granting surface state mediate leakage of electron from edge I to edge II.  Intriguingly, upon incorporating defects into the system, a considerable gain in transmission from edge I to edge II is attained at finite temperatures, implying that both defects and $e-p$ interactions are necessary ingredients for the deterioration of quantized transport in \samplex (Fig. \ref{device}(b,c)).
\subsection{Conclusion}
There are three echelons of dimensionality in transport in MTIs: bulk (3D), surface (2D) and edge (1D) modes, in which the 1D edge states are responsible for the QAHE.  Lack of quantized transport either proceeds from destruction of the edge modes, and/or leakage of current from the bulk and/or surface.  Observation of a topological non-trivial gap above sub-Kelvin temperature rules out the possibility of annihilation of chiral edge states being the culprit, which also has been demonstrated by prior transport studies \cite{Yasuda2020, PhysRevLett.115.057206, Fijalkowski2021}.  Moreover, since STS measurements indicate that the bulk states are energetically well separated from the edge states, such circumstance is unlikely as well. Thus, our detailed STS studies narrow down the perpetrator to be the leakage within the surface states. 

We have revealed in this work that there are spatial modulations in surface conduction and valence bands on the meV-scale. Given that the two bands are only separated by $\approx 13.5$ meV (at 4.2 K), slight spatial modulations in bands can give rise to regions in which the valence band maximum is higher than the conduction band minimum at a different point separated by $<3$ nm distance within the sample.  The presence of energy misalignment in $\Delta_\mathrm{s}$ is a facet of both (1) disorder particularly owing to Bi and Cr dopants in the Sb$_2$Te$_3$ system and (2) electron-phonon interaction of the surface states. Numerical calculations presented here indicate that although increasing the number of Bi- and Cr-dopants alone result in an increase in transmission of electrons across the spatially modulated surface bands, there is a further surge when incorporating the effects electron-phonon interactions. Such detrimental effect of $e-p$ interactions emanates from its renormalization effect to the surface band such that the magnitude of $\Delta_\mathrm{s}$ is reduced at finite temperatures. Consequently, the effects of disorder-induced spatial band modulations intensify, yielding a higher degree of leaky current to tunnel across the gap. Notably, the effects of phonons alone are not strong enough to generate an effective gap-less state without the role of disorder. Only when combining both effects together, an effectively gap-less phase is created, which results in leakage currents throughout the surface state along with the existing edge states, thus shrouding the QAHE. Although we have numerically simulated the effects of Bi- and Cr-dopants, we can expect the same for V-dopants, as they also induce a shift in the local chemical potential.

In other words, suppressing either one of disorder or $e-p$ interactions would allow for QAHE at elevated temperatures. Unfortunately, owing to the futility of attempting to engineer $e-p$ interactions for a known crystalline structure, a method of growing samples with a spatially highly uniform Dirac point and a perfect alignment of the Fermi energy with the Dirac point will be necessary to align the topological surface gap spatially so as to improve the quantum transport properties of MTIs, at least for the \samplex system. With the recent development of the capability of an intrinsic antiferromagnetic topological insulator, MnBi$_2$Te$_4$, to exhibit quantized transport at a temperature of 1.4 K at zero field in a 5 septuple layer sample without the need of any doping \cite{doi:10.1126/science.aax8156}, moreover with a larger surface gap of 60 meV - 85 meV \cite{Otrokov2019, PhysRevResearch.1.012011}, MnBi$_2$Te$_4$ can be considered as a compelling platform. Yet, up to date, QAHE in these materials have been limited to exfoliated nanoflakes due to the lack of capability in growing well-engineered MBE thin films. Efforts towards synthesizing thin films of MnBi$_2$Te$_4$ that exhibit quantized transport at elevated temperatures are still needed towards realistic technological applications based on the QAHE \cite{LIU2022126677, nano11123322, doi:10.1021/acs.nanolett.1c02493}.

\subsection{Acknowledgments}
The authors gratefully acknowledge joint support for this work by the Army Research Office under the MURI program (US) (Award $\#$W911NF-16-1-0472), National Science Foundation (US) (Award No. 1733907) under the Physics Frontier Centers program for the Institute for Quantum Information and Matter (IQIM) at the California Institute of Technology, and the Kavli Foundation. N.-C.Y. acknowledges partial support from the Thomas W. Hogan Professorship at Caltech, the Yushan Fellowship awarded by the Ministry of Education in Taiwan, and the Yushan Fellow Distinguished Professorship at the National Taiwan Normal University (TW) in Taiwan.


\providecommand{\noopsort}[1]{}\providecommand{\singleletter}[1]{#1}%

\end{document}